

\input epsf.tex


\headline={\ifnum\pageno=1\firstheadline\else
\ifodd\pageno\rightheadline \else\leftheadline\fi\fi}
\def\firstheadline{\hfil}
\def\rightheadline{\hfil}
\def\leftheadline{\hfil}
        \footline={\ifnum\pageno=1\firstfootline\else\otherfootline\fi}
\def\firstfootline{\rm\hss\folio\hss}
\def\otherfootline{\hfil}

\font\twelvebf=cmbx10 scaled\magstep 1
\font\twelverm=cmr10 scaled\magstep 1
\font\twelveit=cmti10 scaled\magstep 1

\font\tenrm=cmr10

\parindent=1.5pc
\hsize=6.0truein
\vsize=8.5truein
\nopagenumbers
\def\scrip{{\cal I}^+}
\def\scrim{{\cal I}^-}

\vglue 1cm
\centerline{\twelvebf BLACK HOLE ENTROPY AND THE SEMICLASSICAL
APPROXIMATION${}^*$}
\vglue 1.6cm
\centerline{\tenrm SAMIR D. MATHUR}
\baselineskip=26pt
\centerline{\twelveit Center for Theoretical Physics }
\baselineskip=16pt
\centerline{\twelveit Massachussetts Institute of Technology, Cambridge, MA
02139, USA}
\vglue 2.8cm
\centerline{\tenrm ABSTRACT}
\vglue 1.3cm
{\rightskip=3pc
 \leftskip=3pc
 \twelverm\baselineskip=15pt\noindent
We compute the entropy of the Hawking radiation for an evaporating black hole,
in 1+1 dimensions and in 3+1 dimensions. We investigate the validity of the
semiclassical approximation for
the evaporation process. It appears that there might be  a
large entropy of entanglement between the classical degrees of freedom
describing the black hole and the radiation fields
when the theory of quantum gravity plus matter is considered.
\vglue 0.6cm}

\vfil
 \tenrm\baselineskip=12pt\noindent
${}^*$\quad Invited talk given at the International Colloquium on Modern
Quantum Field Theory II
at TIFR (Bombay) January 1994. (Preprint \#MIT-CTP-2304)
\eject
\twelverm
\baselineskip=14pt

\leftline{\twelvebf 1. Introduction}
\vglue 0.4cm
Recently there has been a renewed interest in the black hole evaporation
problem, and the associated problem of information loss.  One reason is the
construction of 1+1 dimensional models where evaporating holes can be easily
studied${}^{1,2}$. Another reason is  a spate of work on the proposal of 't
Hooft${}^3$ that the black hole evaporation process may not be semiclassical.
This idea is based on the fact that although the Hawking radiation at $\scrip$
is low frequency ($\sim M^{-1}$) it originates in very high frequency vacuum
modes at $\scrim$, the latter frequency being $\sim e^M$ times the planck
frequency. ($M$ is the mass of the black hole in planck units.)

The need to consider super-planckian frequencies suggests that the black hole
evaporation process involves quantum gravity, and cannot be approximated by a
semiclassical calculation of `field theory on curved space'. Susskind et. al.
have argued that the information of the infalling matter is transferred at the
horizon to the Hawking radiation, thus avoiding information loss${}^4$. The
paradox then is: How does this information transfer occur when seen from the
frame of an infalling  observer, who probably sees nothing special at the
horizon? For this issue Susskind suggests a breakdown of Lorentz symmetry at
large boosts, and a principle of `complementarity' which says that one can
observe either the state outside the horizon  or the state inside, but somehow
it should make no sense to talk of both  states at at the same time${}^5$.

The strongest support for such a conjecture of complementarity comes from the
work of ref. 6. The study of quantum gravity plus matter in 1+1 dimensions
reveals large commutators between operators representing infalling matter and
operators of the radiating fields at $\scrip$. This result emphasises the role
of quantum gravity in the evaporation process, and suggests a transfer of
information from the infalling matter to the Hawking radiation.

Can we conclude  that  black hole evaporation is not a semiclassical process?
There seems to be no universal agreement on this point. One reason seems to be
that there is no sufficiently explicit identification of what goes wrong with
the semiclassical reasoning.  A second  reason is that different quantities are
computed by proponents of different views on the information question; this
obscures the precise role of quantum gravity in the problem. A third, more
minor, reason is the puzzlement over the role of certain boundary conditions
assumed in ref. 6.

In this talk we propose an approach that should eliminate the above
difficulties. Issues of information and entropy are best examined through the
`state' of quantum fields on a spacelike hypersurface, rather than through
correlation functions of the quantum theory. Consider a hypersurface such as
$\Sigma$ shown in figure 1. This is a `1-geometry' in 1+1 spacetime, and a
`3-geometry' in 3+1 spacetime. The matter on this slice we separate into three
classes. At the extreme left we have the infalling matter that makes the black
hole; we call this component `A'. Next we have the quantum radiation that falls
into the singuarity (`B'). At the extreme right we have quantum radiation (`C')
which escapes to $\scrip$ as Hawking radiation.

\bigskip
\centerline{\epsfysize=5.3 in \epsffile{bfig.eps}}
\bigskip
\tenrm\baselineskip=12pt\noindent
Figure 1. The semiclassical geometry of a black hole in the RST model.
\twelverm
\baselineskip=14pt
\bigskip

In the first part of the talk we examine the semiclassical approach. Matter `A'
is classical, with mass $M$, and the metric is determined by $M$ and
$<T_{\mu\nu}>$  from the radiation. On this given 2-dimensional spacetime
geometry the quantum field `B' and `C' are examined, and the entropy of
entanglement between these components computed${}^7$. The same redshift that
yields the thermal nature of Hawking radiation also yields, in this
calculation, the entropy of the Hawking radiation. Thus we see more
transparently the origin of `black hole thermodynamics'. (If evaporation were
ignored in the black hole geometry, then the Hawking  temperature could be
calculated but the resulting entropy would be infinite.)

In the second part of the talk we propose a criterion for the validity of the
semiclassical approximation in the evaporation process. The state of the
quantum gravity - matter system is described by a wavefunction in `extended
superspace'. But superspace describes $D-1$ dimensional spacelike
hypersurfaces, not $D$ dimensional spacetime. An approximate $D$ dimensional
spacetime emerges by a WKB approximation on the wavefunction, which is given to
satisfy the Wheeler-de Witt equation in extended superspace. Simple estimates
suggest the the semiclassical approximation breaks down at hypersurfaces
`crooked enough' to capture both the Hawking radiation and the infalling
matter.


\vglue 0.6cm
\leftline{\twelvebf 2. Entropy in the Semiclassical Approximation}
\vglue 0.4cm
Consider first the RST model of dilaton gravity coupled to massless scalar
fields${}^2$. The matter component `A' defined above is classical. The
entanglement of `B' and `C' is the entropy of pair creation: one member of the
pair falls into the singularity while the other escapes to $\scrip$.

Let $\sigma^-$ be the null Minkowski co-ordinate at $\scrip$, and $y^+$ the
null Minkowski co-ordinate at $\scrim$. We wish to compute the entropy
collected by an observer at $\scrip$, who collects radiation upto some time
$\sigma_1^- < 0$. (The radiation finishes at $\sigma^-=0$ because all   of
the mass $M$ has evaporated away.)

In our region of interest a null ray at $\sigma^-$ can be followed back from
$\scrip$, reflected off the `strong coupling boundary'  $\Omega=\Omega_{\rm
crit}$ and followed back to $\scrim$ to reach at some  co-ordinate $y^+$.
$\sigma^-(y^+)$ is given by
$$\sigma^-~=~-{1\over \lambda}\ln [ {\lambda x^-_s e^{_\lambda y^+}+{\pi M\over
\lambda\kappa}\over \lambda x_s^-+{\pi M\over \lambda\kappa}}]\eqno{(1)}$$
Here $x_s^-={\pi M\over \kappa \lambda^2}(1-e^{-4\pi M/ \kappa \lambda})^{-1}$,
$\lambda$ is the cosmological constant (which sets the planck scale) and
$\kappa=N/12$, $N$ being the number of matter fields.

For $y^+\rightarrow 0^-$ (1) can be approximated as
$$\sigma^-~=~-{4\pi M\over \kappa \lambda^2}~-~{1\over \lambda}\ln(-\lambda
y^+)\eqno{(2)}$$
The Bogoliubov transformation given by (2) converts the vacuum at $\scrim$ to a
thermal state at $\scrip$ with temperature $T={\lambda\over 2\pi}$.

To compute entropy we recall a result of Srednicki${}^8$.  Consider a free
scalar field on a
1-dimensional lattice, with lattice spacing $a$. Let this field be in the
vacuum state. Select a region of length $R$ of this lattice and trace
over the field degrees of freedom outside this region. This gives a
reduced density matrix $\rho$, from which we compute $S= -Tr\{ \rho
\ln \rho \}$  which is the entropy of entanglement  of the selected
region
with the remainder of the lattice.
This entropy is given by
$$ S~=~ \kappa_1 \ln(R/a)~+~\kappa_2\ln(\mu R) \eqno{(3)}$$
for one scalar field. (For $N$ species the result must
be multiplied by $N$.) One finds $\kappa_1=1/6$.
 $\mu$ is an infrared cutoff, and the coefficient $\kappa_2$ is sensitive to
the choice of boundary conditions. (For a detailed discussion of boundary
conditions and a heuristic derivation of the form (3), see ref. 7. For an
alternative analytical derivation of $\kappa_1$ see ref. 9.)

To compute the entropy collected at $\scrip$ in $-\infty<\sigma^-<\sigma_1^-$,
consider the hypersurface $\Sigma$ in fig. 1. $\Sigma$ runs near $\scrip$ upto
$\sigma_1^-$, then bends down to avoid the singularity and reaches
$\Omega=\Omega_{\rm crit}$, remaining spacelike throughout. The instrument
measuring the radiation is  to be switched off at $\sigma_1^-$, but the
switching off process cannot be too sudden, otherwise it will generate extra
radiation which will render the measurement inaccurate.  It seems reasonable to
switch off the radiation over a time $\Delta \sigma^-$ of the order of the
period of the Hawking radiation; the exact choice will not matter.

The entanglement entropy we seek corresponds to separating the field modes on
the left of the switchoff point with the field modes on the right of this
point, with the scale of cutoff being $\sim \Delta \sigma^-$.
The left moving components of the matter fields on $\Sigma$ are not excited,
and with the above choice of $\Delta \sigma^-$ give no significant contribution
to  $S$ in (3).  The right moving components are not in a vacuum state, so we
cannot directly apply (3). But we may use any set of co-ordinates on $\Sigma$
to compute the entanglement of modes. Through any point $P$ on  $\Sigma$,
follow a null ray towards the past until it hits $\Omega=\Omega_{\rm crit}$;
reflect here to a null ray which is followed to $\scrim$, reaching at some
co-ordinate $y^+$. In terms of the co-ordinate $y^+(P)$ the right moving field
modes on $\Sigma$ are in the vacuum state, but the cutoff scale is squeezed to
$a\sim e^{-(4\pi M/ \kappa \lambda+\lambda\sigma_1^-)}\sim \Delta \sigma^-$.
The smallness of the exponential in `$a$' provides the dominant term in (3),
and gives
$$S~\approx~{1\over 12}(4\pi M/ \kappa \lambda+\lambda\sigma_1^-)\eqno{(4)}$$
(The prefactor is ${1\over 12}$ because only the right movers have this
entropy.) When the entire hole has evaporated ($\sigma^-_1=0$)  we have
collected an entropy
$$S~\approx~{4\pi M\over \lambda}~=~2S_{\rm Bek}\eqno{(5)}$$
where we multiplied by $N$ to account for $N$ evaporating fields. $S_{\rm Bek}$
is the Bekenstein entropy of a hole of mass $M$.

We can send in a second shock wave after part of the hole has evaporated, and
compute the total radiated entropy. This radiated entropy can be made
arbitrarily large while keeping  some chosen upper bound for the Bekenstein
entropy during the process. [Similar ideas have  been developed by S. Trivedi
and collaborators${}^{10}$.]

In 3+1 dimensions one would trace out the modes in a ball of radius $R$, with a
cutoff scale `$a$'. The analogue of (3) is${}^8$ $S\approx .30 (R/a)^2$. But
this flat space result will not be relevant for our calculation of the entropy.
The field in 3 space  dimensions may be decomposed into angular modes, and then
each mode satisfies a field equation in the 1+1 dimensional $r,t$ space. The
number of angular modes that are effectively massless and contribute to the
entanglement entropy is $\sim (R/a)^2$ (a large number), and this accounts for
the $R^2$ behavior of the entropy. But in the black hole the first few angular
modes contribute practically all the radiation; the higher modes are scattered
by the potential and cannot be followed from $\scrip$ to $\scrim$ in the
argument used by Hawking${}^{11}$ in establishing the thermal radiation.

Let us assume that only the angular mode $l=0$ is emitted (s-wave
approximation). We do not have an explicit evaporating geometry for the 3+1
hole, but can take as an approximation the relation
$$d\sigma^-~=~-4M(\sigma^-)d(\ln(-y^+)),~~~~M(\sigma^-)~=~({-\sigma^-\over 256
\pi})^{1/3}\eqno{(6)}$$
to replace (1). This approximation gives radiation at  any retarded time
$\sigma^-$ at the temperature appropriate to the remaining mass of the hole;
the mass loss rate is computed for s-wave radiation with one scalar field.
Computing  in the  same manner as for the RST model, we get for the total
entropy
$$S~\approx~8\pi M^2~=~2S_{\rm Bek}\eqno{(7)}$$

{}From a thermodynamic calculation Zurek${}^{12}$ estimated the relation
between $S$ and $S_{\rm Bek}$:
$$R~=~dS/dS_{\rm Bek}~=~{\int_0^\infty dx x^2
\sigma(x)\{xe^x/(e^x-1)-\ln(e^x-1)\}\over \int_0^\infty dx x^3
\sigma(x)/(e^x-1)}\eqno{(8)}$$
Here $ \sigma(\omega)~=~\sum_{l,m} \Gamma_{l,m}(\omega)/[27(\omega M)^2]$ is
the absorptivity of the hole. For the s-wave approximation
$\sigma(\omega)=1/[27(\omega M)^2]$, and then (12) gives $R=2$, in accordance
with our direct computations.

(This part of the work is in collaboration with E. Keski-Vakkuri.)

\vglue 0.6cm
\leftline{\twelvebf 3. Role of Quantum Gravity}
\vglue 0.4cm

As mentioned above, 't Hooft has proposed that quantum gravity would be
important at the horizon because very high frequency matter modes here
contribute to the Hawking radiation that appears at infinity. But the horizon
is a completely `regular' place, with small curvature (and therefore small
gravitational coupling from a field theory point of view). How then can quantum
fluctuations of the metric be important at the horizon?  Below we outline an
approach by which one might indeed see a nonsemiclassicality of gravity in the
black hole, while staying away from the singularity (strong coupling region).

An important feature of black hole formation is that the entire mass $M$ of the
collapsing matter is converted to quantum matter radiation. By contrast, if a
ball of dust collapses to a `star' of radius $R=4M$ (say) then only an order
$\sim 1/M$ of energy is carried away by the excitations of the quantum matter
fields.

Whenever we have particle creation, we have a situation where from a field
theory point of view, the `in-vacuum' $|0>_{\rm in}$ is not equal to the
`out-vacuum'
$|0>_{\rm out}$. In this case we must distinguish between `in-in' amplitudes
and `in-out'  (scattering) amplitudes. This issue for first quantised string
theory was studied in ref [13], where it was argued that incorporating the
backreaction from radiation requires a modification of the path integral to
reflect `in-in' amplitudes rather than scattering amplitudes.

What is the effect of creating a state that evolves to be full of pairs of
`out' particles? One important effect is decoherence. For example it was shown
in [14] that when a sufficient number of particle pairs are produced in a
cosmology, the two time directions contained in the quantum gravity
wavefunction can get decohered from each other. But an interesting situation
arises if we have `too much decoherence'. In this case the wavefunction of
`macroscopic' variables can cease to behave classically [15]. Let us examine
this situation in more detail.

Suppose we have a `classical' degree of freedom $X,~ P_X$. (For example $X$ and
$P_X$ could be the position and momentum of a heavy particle of mass $M$.)
Coupled to $X$ is a `quantum' variable $y,~ p_y$. (This could be a harmonic
oscillator, with frequency dependent on $X$.) The state of the complete system
is $\Psi[X,y]$. If we want a semiclassical description of the system where $X$
is classical and $y$ is quantum, then we must have
$$\Psi[X,y]~\approx~\psi[X]~\psi[y]\eqno{(9)}$$
where $\psi[X]$ is a wavepacket with $X$ {\it and} $P_X$ quite well defined
(say with widths $\Delta X$ and $\Delta P_X$).

But now suppose that after a period of time evolution the state of the variable
$y$ is modified to a highly `squeezed' state, i.e., a state with a lot of
`particle creation'. The precise state so obtained depends in general very
sensitively on the behavior of the variable $X$. The complete state at a later
time can only be written as
$$\Psi[X,y]~=~\sum_i\psi_i[X]~\psi_i[y]\eqno{(10)}$$
Now we lose the  possibility of a semiclassical description, which would have
allowed us to replace the variables $X, ~P_X$ by some classical mean values and
yet describe the variable $y$ through a wavefunction $\psi[y]$. For instance,
let the state to which $y$ evolves depend sensitively on the momentum $P_X$, so
that the change $P_X\rightarrow P_X+\epsilon$ leads to a change
$\psi[y]\rightarrow \psi'[y]$ with $<\psi'y|\psi[y]>\approx 0$. But if this
happens for $\epsilon<<\Delta P_X$ then we cannot factor out a `classical'
looking wavepacket to describe $X$ while $y$ is described by a  state
$\psi[y]$; in other words we cannot obtain a semiclassical description even
though $X$ described a particle with large mass $M$.

Let us now consider the issue of semiclassicality in the black hole.
In quantum gravity the degrees of freedom are `3-geometries' in 3+1 spacetime
(1-geometries in 1+1 spacetime). The 4-dimensional spacetime emerges only by  a
WKB approximation to the wavefunction on the space of 3-geometries. How
accurate is this approximation? It can certainly break down at a singularity.
But it may also break down in a black hole spacetime simply because the `time'
is so different outside and inside the hole. (The problem of time  in the black
hole context has also been discussed by de Alwis${}^{16}$. )

We  consider 1+1 quantum dilaton gravity plus matter. The state in quantum
gravity is described by
a wavefunction on superspace. A point of superspace is a `1-geometry' together
with a configuration of matter fields on this 1-geometry. The `1-geometry'
describes both the metric and the dilaton. At this point one must distinguish
between gauge and non-gauge degrees of freedom for gravity. In discussing the
semiclassical approximation of gravity plus matter, one often considers
explicitly only the gauge degrees; evolving along these gives `time evolution'.
But it is possible to argue on general ground that there is at least one
`non-gauge' degree that will be involved in the description of the wavefunction
of gravity. In ref. 17 this degree is $\cal C$ and its canonical conjugate
$\cal P$; we adopt this notation here as well. The variable $\cal C$ will be a
`classical' variable involved in the description of the gravity variables.  (In
ref. 17 the variable ${\cal C}$ is related to the ADM mass of the hole.)

Consider the  complete wavefunction of gravity plus matter  $\Psi[1-geometry,
f]$. As noted above this  wavefunction on superspace contains both the gauge
degrees of freedom of gravity (which determine the choice of `many fingered
time slice') as well as the dynamical degrees of freedom of the manifold. Let
$\tau$ symbolically indicate the time slice, while ${\cal C}$ indicates the
dynamical degree of freedom which is normally considered to be a `classical'
variable. (We may regard ${\cal C}$ as analogous to the momentum $P_X$ in the
above toy example.)
Thus we may write the wavefunction as $\Psi[\tau,{\cal C},f]$. Here $f$ is the
quantum matter field that will make up the Hawking radiation. (The `classical'
matter (component `A' in the figure) is at this point related to the
`classical' variable ${\cal C}$ of the gravitational field.) $\Psi$ satisfies
the Wheeler-de Wit equation
$$(H_{\rm gravity}~+~H_{\rm
matter})\Psi~=~0\eqno{(11)}$$
We know the state of matter on 1-geometries corresponding to early times; the
quantum field $f$ in particular is in the Gaussian state describing the
`in-vacuum'. From this data the Wheeler - de Wit equation determines $\Psi$
everywhere in superspace.

 For a semiclassical description  where spacetime is classical but $f$ is
quantum we need
$$\Psi[\tau,{\cal C},f]~\approx ~\psi[\tau,{\cal C}]~\psi[\tau,f]\eqno{(12)}$$

The form (11) does appear to be a good approximation for a 1-geometry such as
$\Sigma'$ shown in the figure. But it may not be a good approximation for a
1-geometry such as $\Sigma$, which is  `crooked enough' to capture the Hawking
radiation and the infalling matter, while avoiding the singularity. Here we
expect
$$\Psi[\tau,{\cal C},f]~= ~\sum_{i=1}^N\psi_i[\tau,
C]~\psi_i[\tau,f]\eqno{(13)}$$
This state $\Psi$ has an entanglement between the classical degrees of freedom
describing the hole
with the quantum field $f$. If we insist on replacing the classical degree of
freedom ${\cal C}$ with a mean value $\bar {\cal C}$ then (as in the toy
example done above) we must average over many orthogonal states of the quantum
field $f$, so that we cannot then talk about the quantum state of this field.
In this sense we lose the semiclassical approximation.

 Simple estimates based on the scales involved lead to the conjecture that
$N\sim e^M$ (Here $M$ is the mass of the black hole that will be converted to
radiation.) For the entanglemant between ${\cal C}$ and $f$ we may compute the
reduced density matrix; this will have an entropy
$$S_{\rm q-grav}~=~-{\rm Tr} \{\rho \ln \rho \}~\sim ~ M\eqno{(14)}$$

This `gravitational entropy' is quite different from the entanglement entropy
between the outgoing and infalling Hawking pairs (computed in the previous
section). But because the order of magnitude of the two entropies is
comparable, we see that it makes no sense to localise `information' of the
quantum state of $f$ in the way that one naively expects on a semiclassical
spacetime; one must consider the entanglement of $f$ with the degrees of
freedom determined more nonlocally by the `classical' part of the matter and
gravity degrees of freedom.

[The ideas on decoherence were developed in collaboration with J-G Demers. The
ideas on black holes and quantum gravity were developed in collaboration with
E. Keski-Vakkuri, G. Lifshitz and M. Ortiz.]

\vglue 0.6cm
\leftline{\twelvebf  Acknowledgements}
\vglue 0.4cm

I would like to thank for discussions A. Dabholkar, S.R. Das,  S.P. de Alwis,
J. Goldstone, A. Guth, G. Kunstatter, A. Sen
and S. Trivedi. This work was supported in part by funds provided by the U.S.
Department of Energy (D.O.E.) under
cooperative agreement number $\sharp$DE-FC02-94ER40818.

\vglue 0.6cm
\leftline{\twelvebf  References}
\vglue 0.4cm

\itemitem{1.} C.G. Callan, S.B. Giddings, J.A. Harvey and A.
Strominger, {\twelveit Phys.
Rev.} {\twelvebf D45} (1992) R1005.
\itemitem{2.} J.G. Russo, L. Susskind and L. Thorlacius,
{\twelveit Phys. Rev.} {\twelvebf D46} (1992) 3444; {\twelveit Phys. Rev.}
{\twelvebf D47} (1993) 533.
\itemitem{3.} G 't Hooft, {\twelveit Nucl. Phys.
} {\twelvebf B335} (1990) 138.
\itemitem{4.} L. Susskind, L. Thorlacius, J. Uglum, {\twelveit Phys. Rev.}
{\twelvebf D48} (1993) 3743.
\itemitem{5.} L. Susskind,  {\twelveit Phys.
Rev. Lett.} {\twelvebf 71} (1993) 2367; SU-ITP-93-21 (hep-th 9308139).
\itemitem{6.} E. Verlinde and H. Verlinde, {\twelveit Nucl. Phys.
} {\twelvebf B406} (1993) 43; K. Schoutens, E. Verlinde and H. Verlinde,
{\twelveit Phys.
Rev.} {\twelvebf D48} (1993) R2690; K. Schoutens, H. Verlinde and E. Verlinde,
CERN-TH.7142/94, PUPT-1441.
\itemitem{7.} E. Keski-Vakkuri and S.D. Mathur, MIT-CTP-2272 (hep-th 9312194).
\itemitem{8.} M. Srednicki, {\twelveit Phys. Rev. Lett.} {\twelvebf 71} (1993)
666.
\itemitem{9.}  C. Holzhey, F.
Larsen and F. Wilczek,  PUPT-1454, IASSNS-HEP 93/88 (hep-th 9403108).
\itemitem{10.} T.M. Fiola, J. Preskill, A. Strominger and S.P.  Trivedi,
CALT-68-1918 (hep-th 9403137).
\itemitem{11.} S.W. Hawking, {\twelveit Comm. Math. Phys.} {\twelvebf 43}
(1975) 199; {\twelveit Phys. Rev.} {\twelvebf D14} (1976) 2460.
\itemitem{12.}  W.H. Zurek, {\twelveit Phys. Rev. Lett.} {\twelvebf 49} (1982)
1683.
\itemitem{13.}   S.D. Mathur, MIT-CTP-2275 (hep-th 9306090).
\itemitem{14.}   E. Calzetta and F. D. Mazzitelli, {\twelveit Phys. Rev.}
 {\twelvebf D42} (1990) 4066.
\itemitem{15.}  J.P. Paz and  S. Sinha,
{\twelveit Phys. Rev.} {\twelvebf D44} (1991) 1038.
\itemitem{16.} S.P. de Alwis, {\twelveit Phys.Lett.} {\twelvebf B317} (1993)
46; S.P. de Alwis and D.A. MacIntire, COLO-HEP-333 (gr-qc 9403052).
\itemitem{17.}  D.~Louis-Martinez, J.~Gegenberg and G.~Kunstatter, {\twelveit
Phys.
Lett.} {\twelvebf B321} (1994) 193.

\end